\documentclass[prl,twocolumn,showpcs,amsmath,amssymb]{revtex4}

\usepackage{epsfig}
\usepackage{hhline}
\usepackage{graphicx}
\usepackage{dcolumn}
\usepackage{bm}

\newcommand{\TeV}{\ensuremath{\mathrm{\,Te\kern -0.1em V}}}
\newcommand{\GeV}{\ensuremath{\mathrm{\,Ge\kern -0.1em V}}}
\newcommand{\MeV}{\ensuremath{\mathrm{\,Me\kern -0.1em V}}}
\newcommand{\GeVc}{\ensuremath{\mathrm{\,Ge\kern -0.1em V\!/}c}}
\newcommand{\MeVc}{\ensuremath{\mathrm{\,Me\kern -0.1em V\!/}c}}
\newcommand{\GeVcc}{\ensuremath{\mathrm{\,Ge\kern -0.1em V\!/}c^2}}
\newcommand{\MeVcc}{\ensuremath{\mathrm{\,Me\kern -0.1em V\!/}c^2}}
\newcommand{\Tesla}{\ensuremath{\mathrm{\,T}}}
\newcommand{\ipb}{\ensuremath{\mathrm{\,pb^{-1}}}}
\newcommand{\ips}{\ensuremath{\,\mathrm{ps^{-1}}}}

\newcommand{\um}{\ensuremath{\,\mu\mathrm{m}}}

\newcommand{\sye}[1]{\ensuremath{~\pm #1}}
\newcommand{\ase}[2]{\ensuremath{^{~+ #1}_{~- #2}}}

\newcommand{\dg}{\ensuremath{\mathrm{\Delta\Gamma}}}

\renewcommand{\prl}{\ensuremath{\parallel}}

\newcommand{\jpsi}{\ensuremath{J/\psi}}
\newcommand{\kst}{\ensuremath{K^{\ast0}}}

\newcommand{\vrho}  {\ensuremath{\vec{\rho}}}

\newcommand{\gl}    {\ensuremath{\Gamma_L}}
\newcommand{\gh}    {\ensuremath{\Gamma_H}}

\newcommand{\keV}{\ensuremath{\mathrm{ke\kern -0.1em V}}}

\newcommand{\BdDec}{$B^0_d \rightarrow \jpsi\,\kst$}
\newcommand{\BsDec}{$B^0_s \rightarrow \jpsi\,\phi$}
\begin{document}

\title{
\boldmath
Measurement of the Lifetime Difference Between $B_s$
Mass Eigenstates
\unboldmath
}

\author{\font\eightit=cmti8
\def\r#1{\ignorespaces $^{#1}$}
\hfilneg
\begin{sloppypar}
\noindent 
D.~Acosta,\r {16} J.~Adelman,\r {12} T.~Affolder,\r 9 T.~Akimoto,\r {54}
M.G.~Albrow,\r {15} D.~Ambrose,\r {43} S.~Amerio,\r {42}  
D.~Amidei,\r {33} A.~Anastassov,\r {50} K.~Anikeev,\r {15}
% ...
\end{sloppypar}
\vskip .026in
\begin{center}
(CDF Collaboration)
\end{center}
}
\affiliation{
\vskip .026in
\font\eightit=cmti8
\def\r#1{\ignorespaces $^{#1}$}
\hfilneg
\begin{center}
\r 1  {\eightit Institute of Physics, Academia Sinica, Taipei, Taiwan 11529, 
Republic of China} \\
\r 2  {\eightit Argonne National Laboratory, Argonne, Illinois 60439} \\
\r 3  {\eightit Institut de Fisica d'Altes Energies, Universitat Autonoma
de Barcelona, E-08193, Bellaterra (Barcelona), Spain} \\
\r 4  {\eightit Istituto Nazionale di Fisica Nucleare, University of Bologna,
I-40127 Bologna, Italy} \\
\r 5  {\eightit Brandeis University, Waltham, Massachusetts 02254} \\
\r 6  {\eightit University of California at Davis, Davis, California  95616} \\
\r 7  {\eightit University of California at Los Angeles, Los 
Angeles, California  90024} \\
\r 8  {\eightit University of California at San Diego, La Jolla, California  92093} \\ 
\r 9  {\eightit University of California at Santa Barbara, Santa Barbara, California 
93106} \\ 
\r {10} {\eightit Instituto de Fisica de Cantabria, CSIC-University of Cantabria, 
39005 Santander, Spain} \\
\r {11} {\eightit Carnegie Mellon University, Pittsburgh, PA  15213} \\
\r {12} {\eightit Enrico Fermi Institute, University of Chicago, Chicago, 
Illinois 60637} \\
\r {13}  {\eightit Joint Institute for Nuclear Research, RU-141980 Dubna, Russia}
\\
\r {14} {\eightit Duke University, Durham, North Carolina  27708} \\
\r {15} {\eightit Fermi National Accelerator Laboratory, Batavia, Illinois 
60510} \\
\r {16} {\eightit University of Florida, Gainesville, Florida  32611} \\
\r {17} {\eightit Laboratori Nazionali di Frascati, Istituto Nazionale di Fisica
               Nucleare, I-00044 Frascati, Italy} \\
\r {18} {\eightit University of Geneva, CH-1211 Geneva 4, Switzerland} \\
\r {19} {\eightit Glasgow University, Glasgow G12 8QQ, United Kingdom}\\
\r {20} {\eightit Harvard University, Cambridge, Massachusetts 02138} \\
\r {21} {\eightit The Helsinki Group: Helsinki Institute of Physics and Division of
High Energy Physics, Department of Physical Sciences, University of Helsinki, FIN-00044, Helsinki, Finland}\\
\r {22} {\eightit Hiroshima University, Higashi-Hiroshima 724, Japan} \\
\r {23} {\eightit University of Illinois, Urbana, Illinois 61801} \\
\r {24} {\eightit The Johns Hopkins University, Baltimore, Maryland 21218} \\
\r {25} {\eightit Institut f\"{u}r Experimentelle Kernphysik, 
Universit\"{a}t Karlsruhe, 76128 Karlsruhe, Germany} \\
\r {26} {\eightit High Energy Accelerator Research Organization (KEK), Tsukuba, 
Ibaraki 305, Japan} \\
\r {27} {\eightit Center for High Energy Physics: Kyungpook National
University, Taegu 702-701, Korea; Seoul National University, Seoul 151-742, Korea; and
SungKyunKwan University, Suwon 440-746, Korea} \\
\r {28} {\eightit Ernest Orlando Lawrence Berkeley National Laboratory, 
Berkeley, California 94720} \\
\r {29} {\eightit University of Liverpool, Liverpool L69 7ZE, United Kingdom} \\
\r {30} {\eightit University College London, London WC1E 6BT, United Kingdom} \\
\r {31} {\eightit Massachusetts Institute of Technology, Cambridge,
Massachusetts  02139} \\   
\r {32} {\eightit Institute of Particle Physics: McGill University,
Montr\'{e}al, Canada H3A~2T8; and University of Toronto, Toronto, Canada
M5S~1A7} \\
\r {33} {\eightit University of Michigan, Ann Arbor, Michigan 48109} \\
\r {34} {\eightit Michigan State University, East Lansing, Michigan  48824} \\
\r {35} {\eightit Institution for Theoretical and Experimental Physics, ITEP,
Moscow 117259, Russia} \\
\r {36} {\eightit University of New Mexico, Albuquerque, New Mexico 87131} \\
\r {37} {\eightit Northwestern University, Evanston, Illinois  60208} \\
\r {38} {\eightit The Ohio State University, Columbus, Ohio  43210} \\  
\r {39} {\eightit Okayama University, Okayama 700-8530, Japan}\\  
\r {40} {\eightit Osaka City University, Osaka 588, Japan} \\
\r {41} {\eightit University of Oxford, Oxford OX1 3RH, United Kingdom} \\
\r {42} {\eightit University of Padova, Istituto Nazionale di Fisica 
          Nucleare, Sezione di Padova-Trento, I-35131 Padova, Italy} \\
\r {43} {\eightit University of Pennsylvania, Philadelphia, 
        Pennsylvania 19104} \\   
\r {44} {\eightit Istituto Nazionale di Fisica Nucleare, University and Scuola
               Normale Superiore of Pisa, I-56100 Pisa, Italy} \\
\r {45} {\eightit University of Pittsburgh, Pittsburgh, Pennsylvania 15260} \\
\r {46} {\eightit Purdue University, West Lafayette, Indiana 47907} \\
\r {47} {\eightit University of Rochester, Rochester, New York 14627} \\
\r {48} {\eightit The Rockefeller University, New York, New York 10021} \\
\r {49} {\eightit Istituto Nazionale di Fisica Nucleare, Sezione di Roma 1,
University di Roma ``La Sapienza," I-00185 Roma, Italy}\\
\r {50} {\eightit Rutgers University, Piscataway, New Jersey 08855} \\
\r {51} {\eightit Texas A\&M University, College Station, Texas 77843} \\
\r {52} {\eightit Texas Tech University, Lubbock, Texas 79409} \\
\r {53} {\eightit Istituto Nazionale di Fisica Nucleare, University of Trieste/\
Udine, Italy} \\
\r {54} {\eightit University of Tsukuba, Tsukuba, Ibaraki 305, Japan} \\
\r {55} {\eightit Tufts University, Medford, Massachusetts 02155} \\
\r {56} {\eightit Waseda University, Tokyo 169, Japan} \\
\r {57} {\eightit Wayne State University, Detroit, Michigan  48201} \\
\r {58} {\eightit University of Wisconsin, Madison, Wisconsin 53706} \\
\r {59} {\eightit Yale University, New Haven, Connecticut 06520} \\
{\rm (Received 20 December 2004; published 16 March 2005)} \\
\end{center}
}


\begin{abstract}

We present measurements of the lifetimes and polarization
amplitudes
for \BsDec\ and \BdDec\ decays.
Lifetimes of the
heavy (H) and light (L) mass eigenstates in the $B^0_s$ system
are separately measured for the first time
by determining the relative contributions
of amplitudes with definite $CP$ as a function of the decay time.
Using $203 \pm 15$ $B^0_s$ decays we obtain $\tau_{L} = (1.05
\ase{0.16}{0.13}\sye 0.02$)\,ps and $\tau_{H} = (2.07 \ase{0.58}{0.46} \sye
0.03$)\,ps. 
Expressed in terms of the difference $\Delta\Gamma_s$ and average
$\Gamma_s$, of the decay rates
of the two eigenstates, the results are
$\Delta\Gamma_s/\Gamma_s = (65 \ase{25}{33} \sye 1$)\%,
and $\Delta\Gamma_s = (0.47 \ase{0.19}{0.24} \sye 0.01)\ips$.  \\

\noindent DOI: 10.1103/PhysRevLett.94.101803 \hspace{2.2cm} PACS numbers: 13.25.Hw,11.30.Er,14.40.Nd

\end{abstract}

\maketitle

Particle-antiparticle oscillation occurs for both $B^0_d$ and $B^0_s$
mesons and gives rise, in each system, to two eigenstates with definite
masses (heavy, $m_H$ and light, $m_L$) and widths
($\Gamma_H$ and $\Gamma_L$).  In the
Standard Model (SM), this oscillation is due to second-order
contributions from the weak interaction and depends on the
Cabibbo-Kobayashi-Maskawa (CKM) quark mixing matrix.  Oscillation has been
observed in the $B^0_d$ system, and the mass difference 
($\Delta m \equiv m_H -m_L$) is $\Delta m_d =
(0.507 \pm 0.007) \ips$~\cite{PDG}.  In the $B^0_s$ system, direct observation
of the
oscillation signal has been a challenge: the 95\% CL limit for the
mass difference is $\Delta m_s > 14.4 \ips$~\cite{PDG}.  
The ratio of the decay width difference
$\Delta \Gamma \equiv \Gamma_L - \Gamma_H$ to the average decay width $\Gamma \equiv (\Gamma_L +
\Gamma_H)/2$ is expected to be small (0.2\% - 0.3\%) for
the $B^0_d$ system~\cite{DIGHE02}, 
but sizable for the $B^0_s$ system~\cite{bigi}.
Ref.~\cite{YB} predicts $\Delta \Gamma_s /\Gamma_s = (12 \pm 6)\%$
and Ref.~\cite{rdm2dg} gives the ratio of the decay width difference
to the mass difference.  If $\Delta m_s$ is too large to be directly
measured, a measurement of $\Delta \Gamma_s$ could serve instead,
along with $\Delta m_d$, in tests of the unitarity of the CKM matrix.
In the SM,
the
mass eigenstates in the $B^0_s$ system are expected to be nearly $CP$
eigenstates. The light mass eigenstate is expected to be $CP$-even and
to have a larger decay width, and thus a shorter lifetime, than the
heavy mass eigenstate~\cite{Rosner}. By exploiting decays with known
$CP$ content, it is possible to measure the two decay widths separately.

The decays \BsDec\ and $B^0_d \rightarrow J/\psi K^{\ast 0}(892)$ are
pseudoscalar to vector-vector transitions and are characterized by three
amplitudes.  These amplitudes correspond to transitions in which the
$J/\psi$ and $\phi$ (or $K^{\ast 0}$) have a relative orbital angular
momentum $L$ of 0, 1, or 2.  In the transversity basis~\cite{Rosner},
the decay amplitudes correspond to linear polarization states of the
vector mesons.  The $L = 1$ decays take place via the decay amplitude
$A_\perp$ and correspond to a parity-odd (perpendicular) correlation
between the transverse
linear polarization states of the vector mesons.  The other
two decay amplitudes $A_0$ and $A_\parallel$ lead to decays
corresponding to linear combinations of the parity-even $L = 0$ and $L =
2$ decays.  In this analysis, the fully reconstructed decays
\BsDec\ (with $J/\psi \rightarrow \mu^+ \mu^-$ and
$\phi \rightarrow K^+ K^-$) and \BdDec\ (with $J/\psi \rightarrow \mu^+
\mu^-$ and $K^{*0} \rightarrow K^+ \pi^-$) and their charge conjugates
are used to measure the polarization amplitudes.  The observed final
state particles for the $B^0_s$ and $\overline{B}^0_s$ decays ($\mu^+
\mu^- K^+ K^-$) have a definite $CP$, which depends on $L$, and a
definite angular distribution.  We determine the decay widths for the
heavy and light $B^0_s$ mass eigenstates by measuring the relative
contribution of the $CP$-odd and $CP$-even decays to the observed
angular distribution as a function of the decay time.  The $B^0_d$
decays provide a valuable control sample since they are expected to
occur via similar (parity-odd and parity-even) decay
amplitudes~\cite{Rosner}.

The analysis uses a portion of the data from Run~II at the Fermilab Tevatron
$p\overline{p}$ collider, corresponding to an integrated luminosity of
about $260\ipb$.  The data were collected with the upgraded Collider
Detector at Fermilab (CDF)~\cite{CDFII}, the most relevant components of which are
described below.
A five-layer double-sided silicon microstrip detector, SVX,
provides track measurements at radii between 2.5 and 10.6\,cm and allows
for precise vertex reconstruction.  A cylindrical drift chamber, COT,
with eight alternating axial and stereo superlayers (each super-layer
containing 12 sense wires) provides track
measurements for charged particles between radii of 40
and 137\,cm.  The COT symmetry axis is the main axis of the cylindrical
coordinate system used at CDF.  Both tracking devices are immersed in a
uniform axial $1.4\Tesla$ magnetic field, allowing precision
measurement of the momenta of charged particles in the radial direction
$p_T$.  Planar drift chambers located outside of calorimeters and
additional steel absorbers are used to identify muons.
Muons and charged hadrons are reconstructed
in the central pseudorapidity region $|\eta|<1$.

The three-level trigger system of CDF is used to select events of
interest by requiring two oppositely charged particle tracks, each with
$p_T>1.5\GeVc$ and matched to hits in the muon detector.  About two
million $\jpsi \rightarrow \mu^+\mu^-$ signal candidates were selected
by requiring a reconstructed mass within $80.0\MeVcc$ of the $\jpsi$
mass~\cite{PDG}.  A $B^0_s$ ($B^0_d$) meson candidate is reconstructed
by associating a $\jpsi$ candidate with a pair of tracks, each with
$p_T>0.4\GeVc$, consistent with a $\phi \rightarrow K^+ K^-$ ($\kst
\rightarrow K^+ \pi^-$) decay.  The $K^+K^-$ ($K^+\pi^-$) mass is
required to be within $6.5\MeVcc$ ($50.0\MeVcc$) of the $\phi$ ($\kst$)
pole mass~\cite{PDG}.  When both $K^+\pi^-$ and $\pi^+ K^-$ particle
assignments are kinematically viable in the $\kst$ reconstruction (no
particle identification is used), the one giving the mass closest to the
pole $\kst$ mass is chosen.  This reduces the contribution from
candidates with swapped particle assignment down to about 10\% of the
total signal.  Background is suppressed by requiring $p_T(B^0_{s,d}) >
6.0\GeVc$, $p_T(\phi) > 2.0\GeVc$ and $p_T(\kst) > 2.6\GeVc$. The more
restrictive cut on $p_T(\kst)$ is determined by an optimization
procedure, and is a consequence of the larger combinatorial background
underneath the $\kst$ peak.

We fit the $B$ candidates subject to the constraint that the
four tracks originate from a common point.
In order to improve the mass resolution, the $\mu^+ \mu^-$
mass is constrained to the $\jpsi$ mass.  
To ensure only well measured
vertices, a set of track and vertex quality requirements is
applied~\cite{THESIS1}.  
In particular, all four tracks are required to have
measurements in at least 3 axial layers of
the silicon detector.
The proper decay time, $t$, is determined
from the radial distance $l_T$ 
from the beam axis to the $B$ meson decay vertex,
signed relative to the direction of $\vec{p}_T$ of the $B$
candidate: $ct=c(\vec{l}_T\cdot\vec{p}_T)M_B/p_T^2$.  The position of
the beam axis is
determined using data taken with an inclusive jet
trigger.  The radial profile of the beam is approximately Gaussian with an
RMS of
about $30\um$.


In the $B^0_d$ system, the mass eigenstates are not $CP$ eigenstates,
and the observed decays are flavor-specific with the charge of the $K$
meson identifying whether the decay is that of a $B^0_d$ or a
$\overline{B}^0_d$.  Summing over initially produced $B^0_d$ and
$\overline{B}^0_d$ yields a differential decay rate~\cite{DIGHE99} which
is insensitive to first order to a lifetime difference~\cite{DIGHE02}:
\begin{equation*}
\begin{split}
\frac{d^4\mathcal{P}(\vrho ,t)}{d\vrho \, dt} \propto \,
 & \bigl[ |A_0|^2 \cdot f_1(\vec\rho) + \,
|A_\prl|^2 \cdot f_2(\vec\rho) \\
+ & \, |A_\perp|^2 \cdot f_3(\vec\rho) 
\pm  {\rm Im}(A^*_\prl A_\perp) \cdot f_4(\vec\rho) \\
+ & \, {\rm Re}(A^*_0A_\prl) \cdot f_5(\vec\rho) 
 \pm  {\rm Im}(A^*_0A_\perp)  \cdot f_6(\vec\rho) \bigr]
e^{-\Gamma_d t}.
\end{split}
\end{equation*}
\noindent
Here the upper (lower) sign is used for $K^+\pi^-$($K^-\pi^+$) in the
final state and $\Gamma_d \equiv 1/\tau_{B^0_d}$.  The functions
$f_i(\vec\rho)$ depend on the transversity variables
$\vec\rho\equiv\{\cos\theta,\varphi,\cos\psi\}$, all of which
are defined in Ref.~\cite{DIGHE99}.

In the $B^0_s$ system, the SM expectation is
that $CP$ violation due to mixing is small,
and the mass eigenstates are nearly $CP$
eigenstates. Ignoring $CP$ violation due to mixing, 
and summing the
distributions for initially produced $B^0_s$ and $\overline{B}^0_s$
mesons, 
the interference terms between the $CP$-even and $CP$-odd amplitudes
cancel, leaving:
\begin{equation*}
\begin{split}
\frac{d^4\mathcal{P}(\vrho ,t)}{d\vrho \, dt} \propto \, &
 |A_0|^2 e^{-\gl t} \cdot f_1(\vec\rho)  + \,
|A_\prl|^2 e^{-\gl t} \cdot f_2(\vec\rho) \\  +  & \,
|A_\perp|^2 e^{-\gh t} \cdot f_3(\vec\rho) + \,
{\rm Re}(A^*_0A_\prl) \cdot f_5(\vec\rho)  e^{-\gl t}.
\end{split}
\end{equation*}
In the $B^0_s$ analysis, the observed final
states have definite $CP$ which depends on $L$.
The $CP$-even angular decay terms
($A_{0,\,\prl}$) evolve in time as $e^{-\gl t}$, while the $CP$-odd
angular decay terms ($A_\perp$) evolve as $e^{-\gh t}$.  

For both $B^0_d$ and $B^0_s$ decays, the amplitudes are normalized so
that $|A_0|^2 + |A_\parallel|^2 + |A_\perp|^2 = 1$, and an unobservable
overall phase is removed by setting $\arg(A_0) = 0$.
The decay amplitudes are assumed to be $CP$ conserving.  

An unbinned likelihood fit is performed to mass, $ct$, and $\vrho$
in order to extract the decay amplitudes $A_{0,\parallel,\perp}$ and
decay widths $\Gamma$ ($\Gamma_H$ and $\Gamma_L$ in the case of the $B^0_s$).
Inclusion of the mass information in the fit is crucial for separation
of the signal from the background.  The mass distribution is modeled
with a Gaussian for the signal peak and a linear shape for the
background.  The mass-measurement uncertainty is incorporated for each
candidate.  The probability density function for $ct$ includes positive
exponentials for the signal, a $\delta$-function for the prompt
background (which is about 85\% of the total) and a set of exponentials
for positive and negative decay lengths which describe a short-lived
background component due to mis-measured vertices and a long-lived
contribution due to incorrectly reconstructed heavy-flavor decays.  Each
contribution is convoluted with a Gaussian resolution function, the
width of which is proportional to the uncertainty of the candidate's
$ct$ measurement.  To allow for a systematic underestimate of the
uncertainties, the mass and the $ct$ uncertainties are multiplied by
scale factors determined in the fit.  The $\vrho$
distribution of the signal is parameterized in accordance with the
equations above.  The background distributions in $ct$ and $\vrho$ are
assumed to be uncorrelated.  The latter is described by a shape similar
to that of the signal, but with an independent set of amplitudes.
The relationship of mass, $ct$, and $\vrho$ of the $B^0_d$ candidates
with $K\pi$ mis-assignment to those of correctly reconstructed candidates
is established via Monte Carlo simulation.

Distributions in $\vrho$ are distorted by the detector acceptance, the
trigger efficiency and, most importantly, the kinematic selection
criteria.  We use the method developed for the CDF Run~I measurement of
transversity amplitudes~\cite{THESIS2, cdf00} to account for this
distortion.  With as little as six constants extracted from Monte Carlo
decays generated uniformly in $\vrho$, this method allows one to avoid
the need for explicit parameterization of the distortion in the
likelihood.  All aspects of the fitting are extensively verified using
Monte Carlo simulations.

\begin{figure}[t]
\centering
\epsfig{file=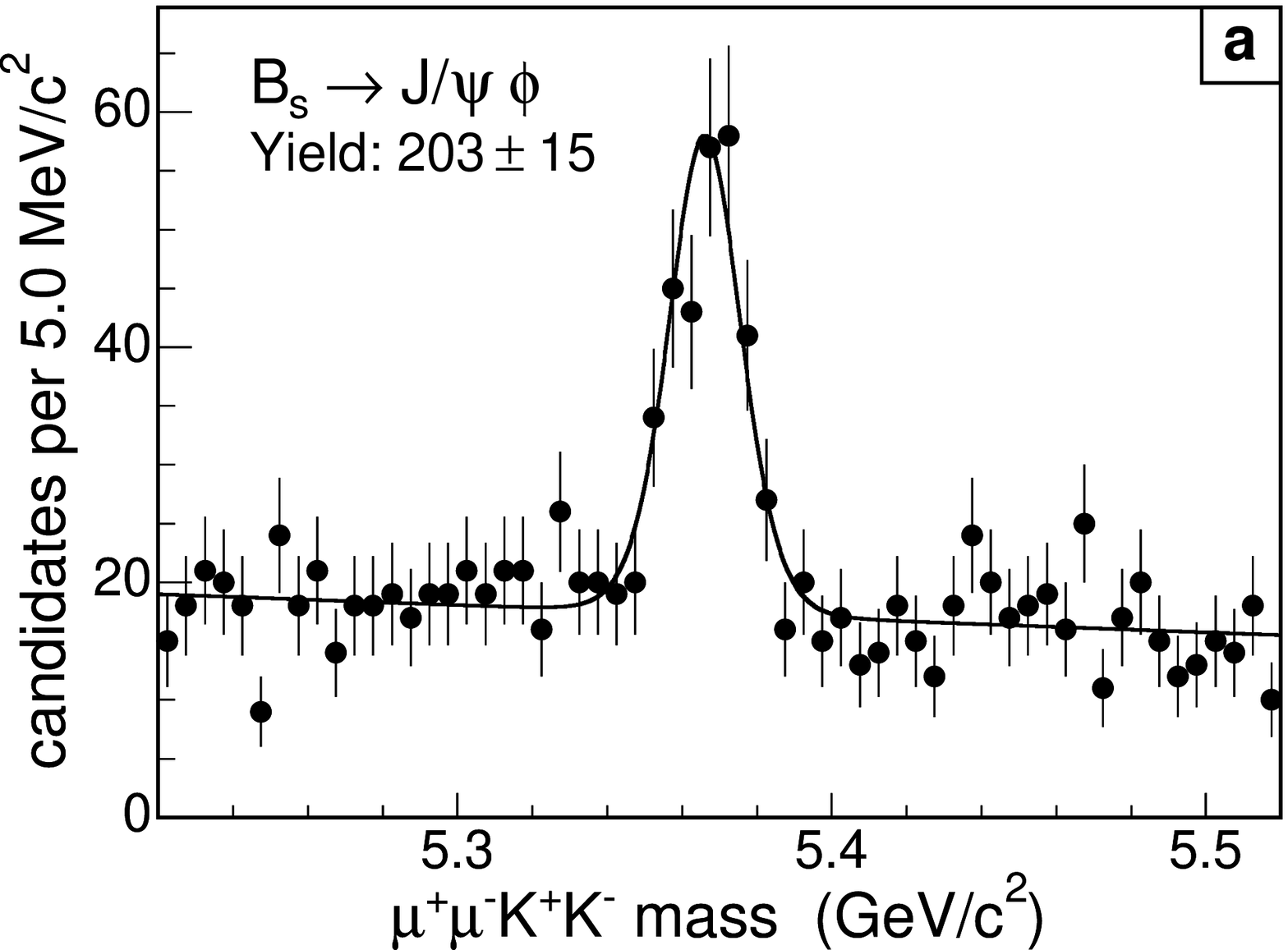, clip=, angle=0, width=1.0\linewidth}
\epsfig{file=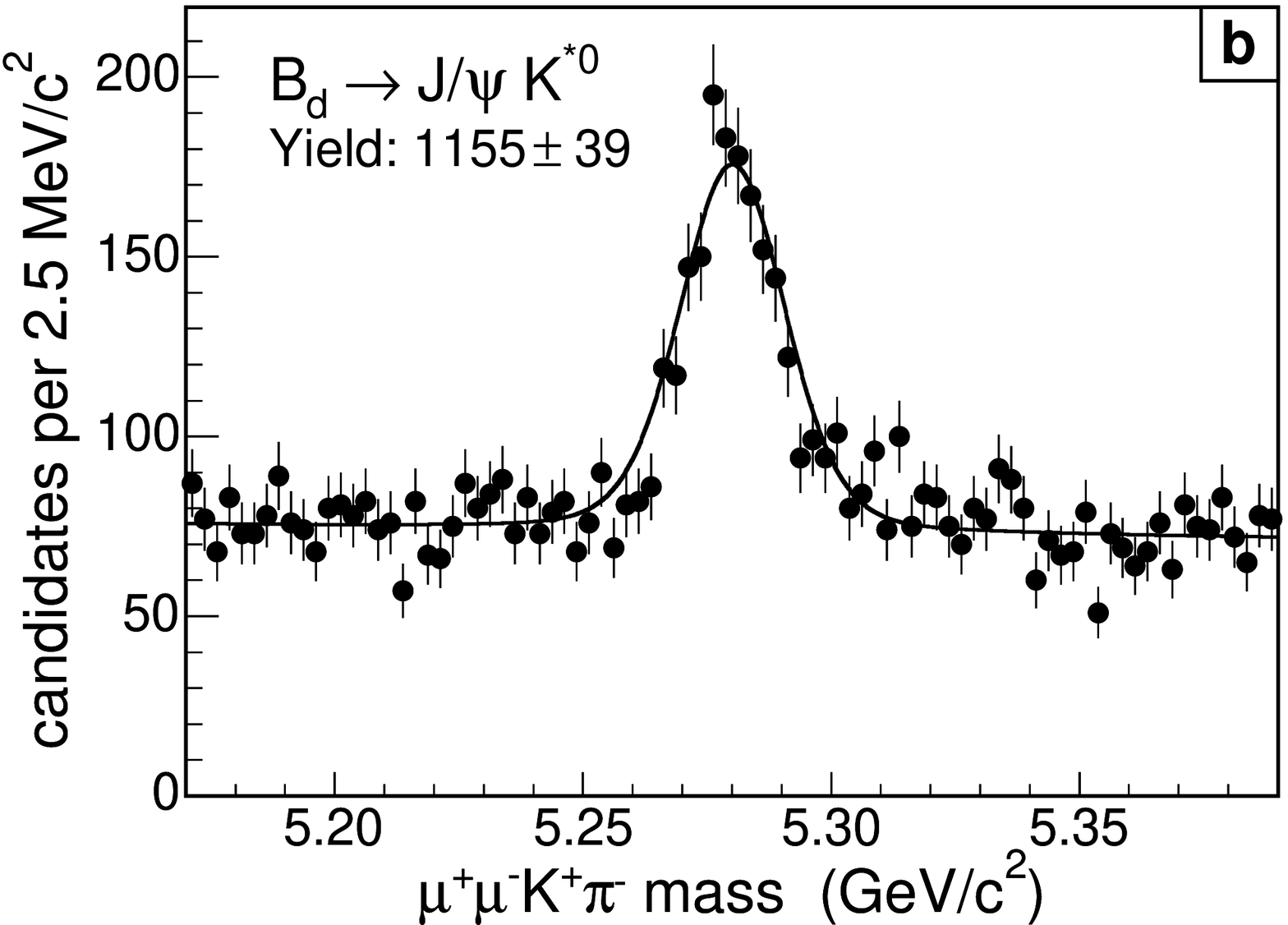, clip=, angle=0, width=1.0\linewidth}
\caption{Mass distribution with the fit projection overlaid:
  (a)~\BsDec\ , (b)~\BdDec\ .}
\label{fig:massfit}
\end{figure}

\begin{figure}[tph]
\centering
\epsfig{file=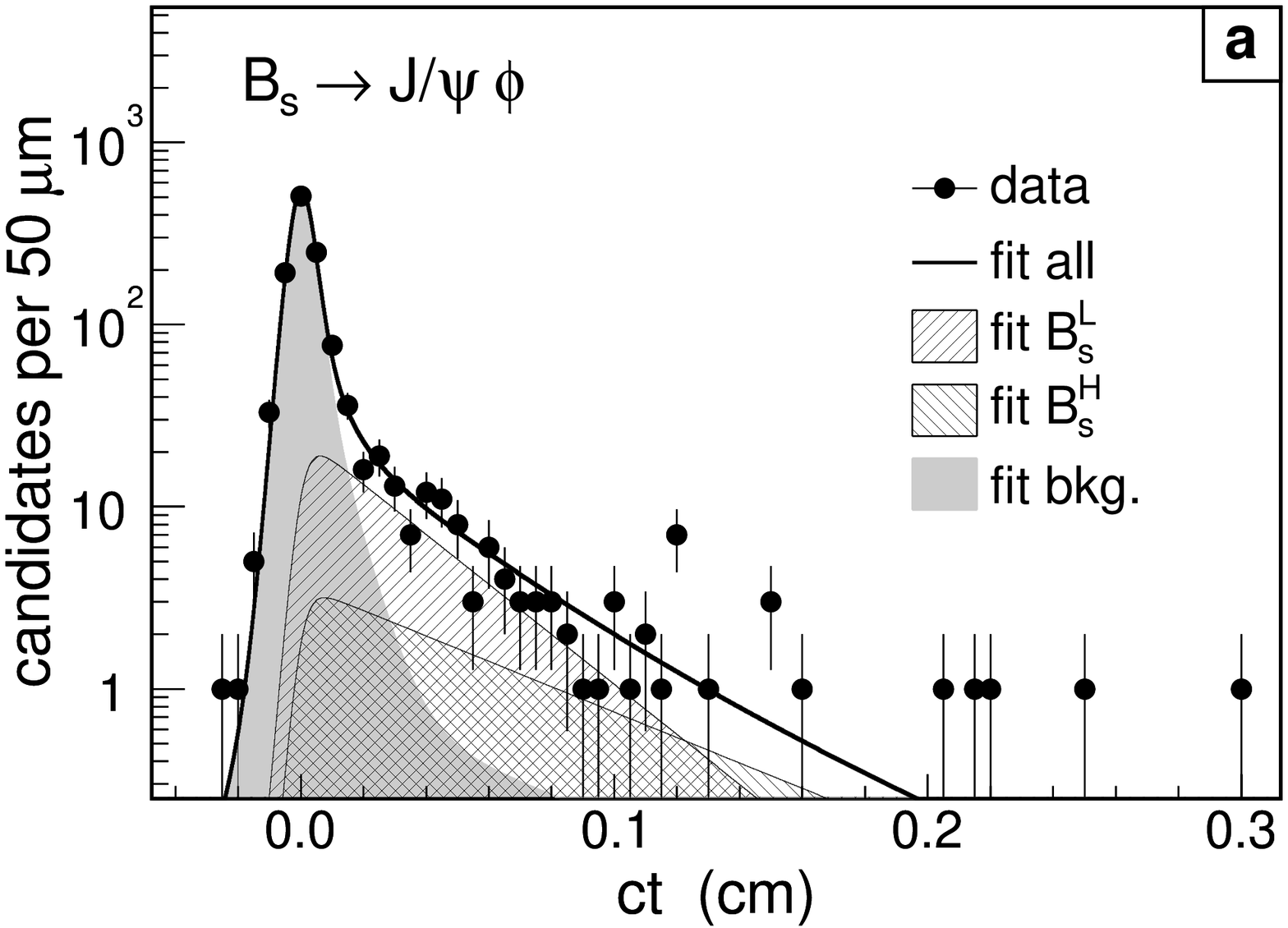, clip=, angle=0, width=1.0\linewidth}
\epsfig{file=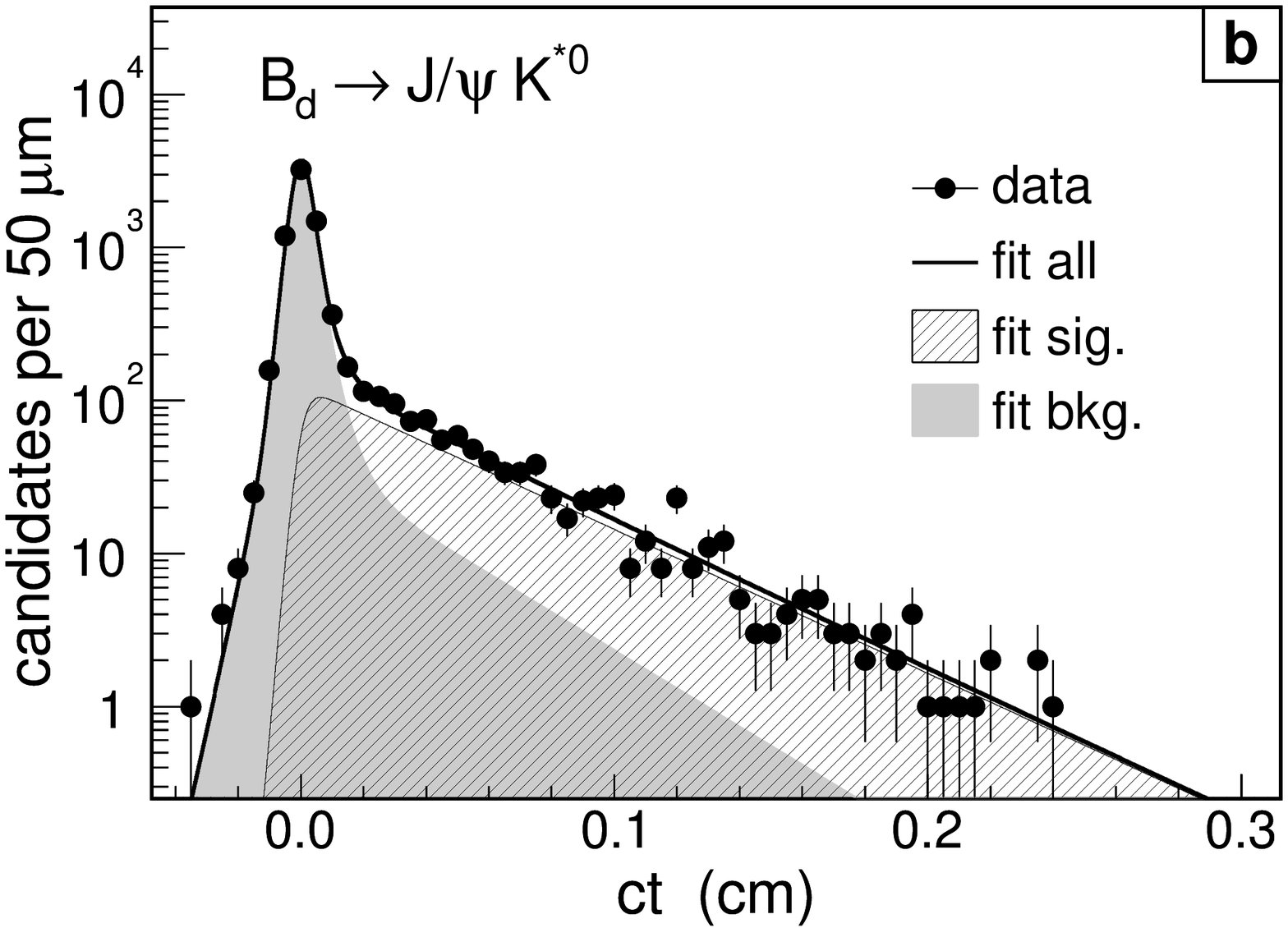, clip=, angle=0, width=1.0\linewidth}
\caption{$ct$ distribution with the fit projection for the
signal and background (bkg.) overlaid:
  (a)~\BsDec\ , (b)~\BdDec\ .}
\label{fig:lifefit}
\end{figure}

\begin{figure*}
\epsfig{file=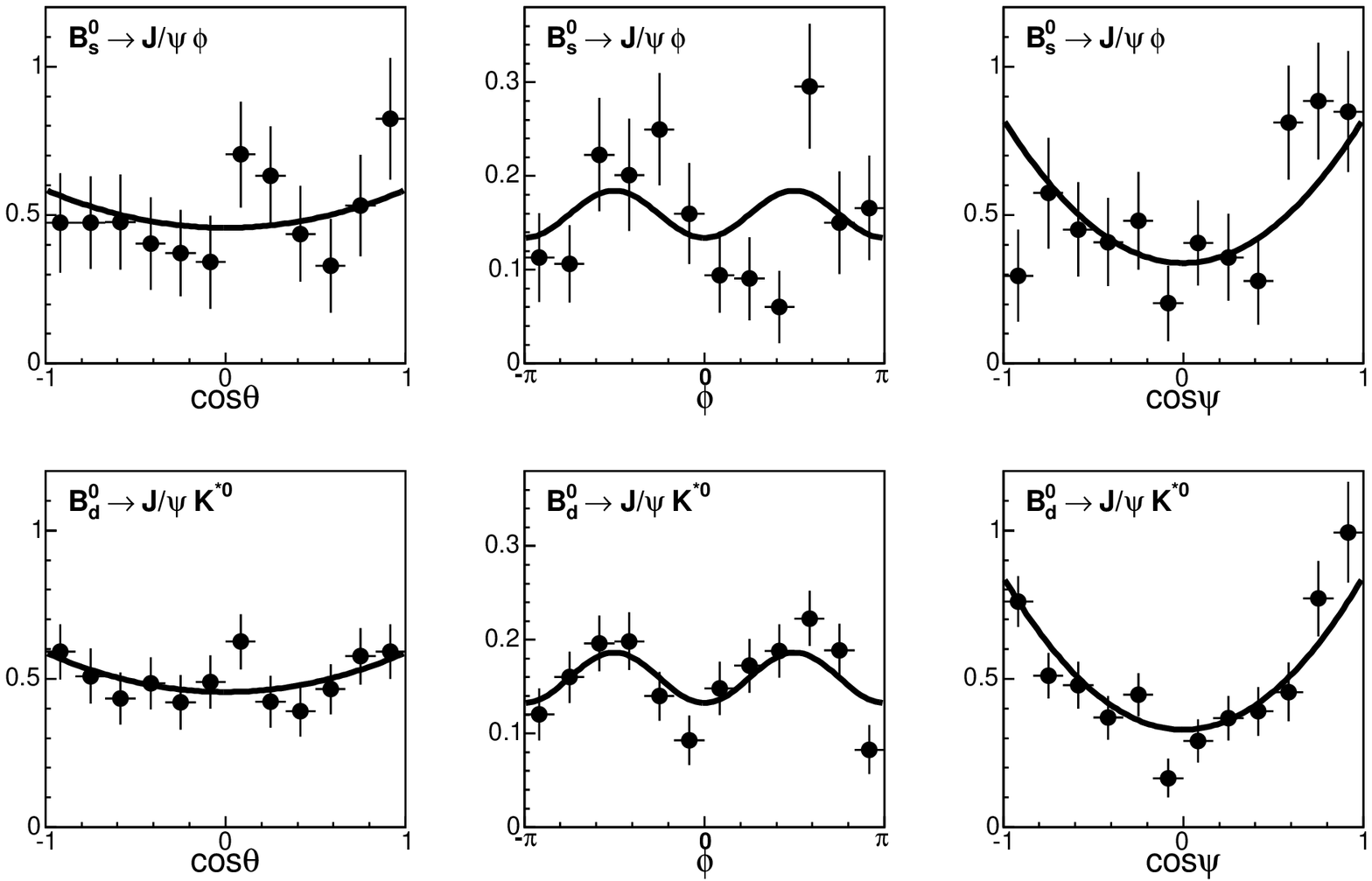, clip=, angle=0, width=0.9\linewidth}
\caption{Projections of the fit onto transversity variables for the
mass-sideband-subtracted acceptance-corrected signal: \BsDec\ (top row),
\BdDec\ (bottom row). A $ct > 0$ cut is applied.}
\label{fig:ang-proj}
\end{figure*}

Data and fit projections in mass and $ct$ for the $B^0_s$ and $B^0_d$ are
shown in Figures~\ref{fig:massfit} and~\ref{fig:lifefit}.  Fits in the
transversity sub-space are illustrated by Figure~\ref{fig:ang-proj}.

The single largest source of systematic uncertainty in the measurement
of the transversity amplitudes of $B^0_{d}$ and $B^0_{s}$ is the choice
of parameterization of the background distribution in $\vrho$.  The
$B^0_s$ transversity amplitudes receive a small contribution to their
systematic uncertainty from a 3.5\% contamination from $B^0_{d}$.  Two
other sources contribute to the uncertainty in $B^0_{d}$ amplitudes: the
way candidates with incorrect $K\pi$ assignment are handled and the
potential contribution of $\mu^+\mu^-K^+\pi^-$ final states that are not
due to $B^0_d\rightarrow\jpsi\kst(892)$ decays, which is estimated to be
less than 4\% of the total signal.  
The systematic
uncertainty in the lifetimes receives contributions from the choice of
the $ct$ parameterization and from 
the SVX alignment.
Slightly larger contributions come from the choice
of the background parameterization in $\vrho$ and, in the case of
$B^0_s$, from $B^0_{d}$ contamination.  For $\dg_s/\Gamma_s$ the only
two sources of systematic uncertainty are from the choice of the
background $\vrho$ parameterization and from $B^0_{d}$ contamination.
Other potential sources of systematic uncertainty, including those from
the method of correcting for distortion of the signal distribution in
$\vrho$ and potential contribution of $B^0_s\rightarrow\jpsi f^0(980)$
decays, were found to be negligible.  The precision of all of the
results of this analysis is statistically limited.

\begin{table}[h]
\caption{Summary of the results of the time-dependent angular analysis
  of \BsDec\ and \BdDec\ decays. A measurement of $\arg(A_\perp)$ is not
  possible for the $B^0_s$ decays because the final state particles do
  not distinguish the decay of a $B^0_s$ from that of a
  $\overline{B}^0_s$.  For the $B^0_s$ decays, any pair of quantities
  describing signal lifetimes ($\tau_{B^0_s} = 1/\Gamma_s$, $\tau_H$,
  $\tau_L$, $\dg_s/\Gamma_s$ and $\dg_s$) may be used as free
  parameters in a fit; separate fits using different pairs are performed
  to obtain directly the results and asymmetric (statistical)
  uncertainties for each of the quantities.}
\centering
\begin{tabular}{@{}c  c  c@{}}
\hline
\hline
               & $B^0_s$ & $B^0_d$ \\
\hline
\vspace{0.2cm}
$N_{sig}$      & $203\pm15$    & $1155\pm39$          \\
$A_0$          & $0.784\pm0.039\pm0.007$ & $0.750\pm0.017\pm0.012$ \\
$|A_\prl|$     & $0.510\pm0.082\pm0.013$ & $0.473\pm0.034\pm0.006$ \\
$|A_\perp|$    & $0.354\pm0.098\pm0.003$ & $0.464\pm0.035\pm0.007$ \\
$\arg(A_\prl)$ & $1.94\pm0.36\pm0.03$    & $2.86\pm0.22\pm0.07$      \\
\vspace{0.2cm}
$\arg(A_\perp)$   &               & $0.15\pm0.15\pm0.04$   \\
$c\tau_{B^0_d}$  &           & $(462\pm15\pm6)$ $\!\um$ \\
$c\tau_{B^0_s}$  &    $(419\ase{45}{38}\pm6)$ $\!\um$        &       \\[0.2em]
$c\tau_{L}$ & $(316\ase{48}{40}\pm6)$ $\!\um$       &   \\[0.2em]
$c\tau_{H}$  & $(622\ase{175}{138}\pm9)$ $\!\um$     &  \\[0.2em]
$\dg_s/\Gamma_s$  & $(65\ase{25}{33}\pm1)$ \%    &    \\[0.2em]
$\dg_s$     & $(0.47\ase{0.19}{0.24}\pm0.01)$ $\!\ips$ \\ 
\hline
\hline
\end{tabular}
\label{tab:fitresults}
\end{table}

Results from the time-dependent angular analysis are given in
Table~\ref{tab:fitresults}. The
$B^0_d$ decay amplitudes and phases are of comparable precision and in
agreement with results from BaBar~\cite{babar01} and
Belle~\cite{belle02}, and the lifetime is in agreement with the world average
value~\cite{PDG}.  
Previous results~\cite{cdf00}
for the $B^0_s$ decay amplitudes, obtained from a time-integrated
analysis, are in agreement with the results obtained in this analysis.
Within uncertainties, the amplitudes for the $B^0_s$ and $B^0_d$
decays are in agreement, as is expected in the limit of
$SU(3)$ flavor symmetry~\cite{DIGHE99}. Explicitly requiring
exact $SU(3)$ symmetry by setting the $B^0_s$ decay amplitudes
to be equal to those of the $B^0_d$, gives a consistent result for
$\Delta \Gamma_s/\Gamma_s$ within uncertainties.

It is predicted~\cite{bigi} that the $B^0_d$ and $B^0_s$ total decay
widths should be equal to within 1\%.  This expectation can be used as a
constraint in the $B^0_s$ fit by requiring
$1/\Gamma_s\equiv\tau_{B^0_s}\equiv2\tau_H\tau_L/(\tau_H+\tau_L) =
\tau_{B^0_d}\equiv1/\Gamma_d$, with $c \tau_{B^0_d} = 460.8 \pm 4.2\um$,
the known value for the $B^0_d$ lifetime~\cite{PDG} with an additional
1\% uncertainty added in quadrature.  By applying this constraint in the
fit, we find $\Delta \Gamma_s/\Gamma_s = (71 \ase{24}{28} \pm 1)$\% and
$\Delta \Gamma_s = (0.46 \ase{0.17}{0.18} \pm 0.01)$ ps$^{-1}$.
Although the uncertainties are still sizable, the fits with and without
the $\Gamma_s=\Gamma_d$ constraint favor a non-zero value for
$\dg_s/\Gamma_s$.

Monte Carlo methods are employed to estimate the probability for an
experiment with similar statistical sensitivity to yield
$\dg_s/\Gamma_s$ as large as is observed in this analysis.  For the SM
expectation, $\dg_s/\Gamma_s = 12\%$, one experiment in 84 (204) would
give a result larger than that obtained from the unconstrained
(constrained) fit.  If no lifetime difference were expected,
$\dg_s/\Gamma_s = 0$, one experiment in 315 (718) would give a result
larger than that obtained from the unconstrained (constrained) fit.

A lifetime difference should result in an increase of the fraction of
$CP$-odd \BsDec\ decays obtained in a time-integrated angular fit as a
function of a cut on $ct$.  The time-integrated $CP$-odd fraction
extracted from fits for four values of the $ct$ cut are shown in the
second column of Table~\ref{tab:cpodd-vs-ct}.  Using $c\tau_L$ and
$c\tau_H$ obtained from the time-dependent fit and the observation that
the time integrated $CP$-odd fraction is 20\% for $ct > 0$, one obtains
expected fractions for the other $ct$ cuts, shown in the third column in
Table~\ref{tab:cpodd-vs-ct}, which are in agreement with the
observations.  The \BdDec\ decays provide an important cross check of
the results obtained for the \BsDec\ decays.  Fits for the
time-integrated fraction of parity-odd $B^0_d$ decays show, as expected,
that the fraction remains unchanged with respect to cuts on $ct$ (last
column of Table~\ref{tab:cpodd-vs-ct}).  In addition,
a fit of the $B^0_d$ data
can be performed allowing two lifetime components.  
The results are consistent with no lifetime
difference for the full sample ($\dg/\Gamma = (15 \pm 12)\%$),
as well as for independent subsamples having a
statistical sensitivity similar to the $B^0_s$ decay sample.
This result is not a measurement of a lifetime
difference in the $B^0_d$ system, but rather a cross check of
the analysis technique.
\begin{table}[h]
\centering
\caption{Time-integrated $CP$-odd $B^0_s$ and parity-odd $B^0_d$ fractions 
(in \%) vs. a
cut on the decay length, $ct$.}
\begin{tabular}{l  l  c  c }
\hline
\hline
$ct$ cut  &\;\;\;$B^0_s$ fitted  &\;\;\;$B^0_s$ expected  &\;\;\;$B^0_d$ fitted \\
\hline
$>  0\um\phantom{00}$       &\;\;\;$20\pm9$  &\;\;\;$20$ (reference) &\;\;\;$22\pm4$\\
$>150\um$                   &\;\;\;$24\pm10$  &\;\;\;$24$      &\;\;\;$23\pm4$\\
$>300\um$                   &\;\;\;$30\pm13$  &\;\;\;$29$      &\;\;\;$23\pm4$\\
$>450\um$                   &\;\;\;$39\pm12$  &\;\;\;$34$      &\;\;\;$24\pm5$ \\
\hline
\hline
\end{tabular}
\label{tab:cpodd-vs-ct}
\end{table}

In conclusion, we have performed the first time-dependent angular
analysis of \BsDec\ decays and have performed a similar measurement
with \BdDec\ decays. The
measured $B^0_d$ polarization amplitudes
are of comparable precision to, and in agreement
with, previously published results.
The measured $B^0_s$ polarization amplitudes
are the most precise available.  Analysis of the \BsDec\ decays
indicates a non-zero lifetime difference between
the heavy and light mass eigenstates of
the $B^0_s$ system.  The result obtained, 
$\Delta \Gamma_s/\Gamma_s = (65 \ase{25}{33} \pm 1)\%$, 
has a central value
larger than the SM expectation
of $(12 \pm 6)\%$.


We thank the Fermilab staff and the technical staffs of the
participating institutions for their vital contributions. This work was
supported by the U.S. Department of Energy and National Science
Foundation; the Italian Istituto Nazionale di Fisica Nucleare; the
Ministry of Education, Culture, Sports, Science and Technology of Japan;
the Natural Sciences and Engineering Research Council of Canada; the
National Science Council of the Republic of China; the Swiss National
Science Foundation; the A.P. Sloan Foundation; the Bundesministerium
fuer Bildung und Forschung, Germany; the Korean Science and Engineering
Foundation and the Korean Research Foundation; the Particle Physics and
Astronomy Research Council and the Royal Society, UK; the Russian
Foundation for Basic Research; the Comision Interministerial de Ciencia
y Tecnologia, Spain; and in part by the European Community's Human
Potential Programme under contract HPRN-CT-20002, Probe for New Physics.

\end{document}